**Micromagnetic Simulation of Amorphous Ferrimagnetic TbFeCo Films with Exchange Coupled Nanophases**

Chung T. Ma, Xiaopu Li, and S. Joseph Poon

Department of Physics, University of Virginia, Charlottesville, Virginia 22904, USA

## Abstract

Amorphous ferrimagnetic TbFeCo thin films are found to exhibit exchange bias effect near the compensation temperature by magnetic hysteresis loop measurement. The observed exchange anisotropy is believed to originate from the exchange interaction between the two nanoscale amorphous phases distributed within the films. Here, we present a computational model of phase-separated TbFeCo using micromagnetic simulation. Two types of cells with different Tb concentration are distributed within the simulated space to obtain a heterogeneous structure consisting of two nanoscale amorphous phases. Each cell contains separated Tb and FeCo components, forming two antiferromagnetically coupled sublattices. Using this model, we are able to show the existence of exchange bias effect, and the shift in hysteresis loops is in agreement with experiment. The micromagnetic model developed herein for a heterogeneous magnetic material may also account for some recent measurements of exchange bias effect in crystalline films.

## 1. Introduction

Amorphous ferrimagnetic (FiM) rare earth (RE) transitional metal (TM) thin films have been widely studied for its applications in high-density low-current spintronic devices and ultrafast magnetic switching [1-4]. Recently, all-optical switching using ultrafast lasers in RE-TM thin films have been investigated [5,6]. There are several advantages of amorphous FiM RE-TM thin films. For example, TbFeCo thin films have strong perpendicular magnetic anisotropy (PMA) and can be synthesized at room temperature requiring no epitaxial growths [7]. In this material, RE and TM form two ferromagnetic (FM) sublattices, which couple antiferromagnetially. Magnetic properties such as magnetization and coercivity are largely influenced by the compensation temperature ($T_{\mathrm{comp}}$), which can be tuned by varying the composition [8]. In a recent paper, we reported the existence of exchange bias (EB) effect in amorphous TbFeCo thin films near the compensation temperature [9]. Two nanoscale amorphous phases were observed to coexist in the amorphous films using atom probe tomography, scanning transmission electron microscopy, and energy dispersive spectroscopy. This exchange anisotropy was proposed to originate from the exchange interaction of these two distributed nanoscale phases with different Tb concentrations. Magnetic modeling is needed to confirm the origin of this EB effect and furthermore, to investigate the size effect of the nanoscale phases in this kind of heterogeneous magnetic materials.

FiM materials have been extensively studied in numerical modeling. Various methods, ranging from Monte Carlo to mircomagnetics, have been employed to study the temperature dependence of magnetic properties in FiM [10-12]. Magnetic reversal dynamics and EB effects involving FiM have also been investigated numerically [13-20]. Monte Carlo Metropolis sampling was employed to investigate the EB effect in FM core/FiM shell structure [16]. Using micromagnetic model, EB effects were obtained and compared to experimental results in exchange-coupled FiM/FM heterostructures [17,18], FiM bilayers [19], and multilayers [11,12,20]. In this paper, we present a micromagnetic model to study heterogeneous magnetic materials with two interpenetrating nanoscale phases. We start by adopting the micromagnetic model to represent the two sublattices, where each sublattice evolves under its own Landau-Lifshitz-Gilbert (LLG) equation. We apply this model to FiM/FiM heterostructure by introducing two types of cells that are distributed throughout the modeling space, representing the two nanoscale phases observed in experiment [9]. Using this model that describes an interpenetrating heterogeneous system, the calculated EB effects are compared to the experimental results.

## 2. Formulation

### *2.1 The two-sublattice model*

In the two-sublattice model, each cell contains separated Tb and FeCo components, forming two antiferromagnetically coupled sublattices. Following Mansuripur [13], we allow each component evolves under LLG equation.

$$\dot{\boldsymbol{M}}_{\mathrm{Tb}} = -\gamma\left(\boldsymbol{M}_{\mathrm{Tb}} \times \boldsymbol{H}_{\mathrm{eff}_{\mathrm{Tb}}}\right) + \frac{\alpha}{M_{\mathrm{Tb}}}\left(\boldsymbol{M}_{\mathrm{Tb}} \times \dot{\boldsymbol{M}}_{\mathrm{Tb}}\right) \tag{1}$$

$$\dot{\boldsymbol{M}}_{\mathrm{Fe}} = -\gamma\left(\boldsymbol{M}_{\mathrm{Fe}} \times \boldsymbol{H}_{\mathrm{eff}_{\mathrm{Fe}}}\right) + \frac{\alpha}{M_{\mathrm{Fe}}}\left(\boldsymbol{M}_{\mathrm{Fe}} \times \dot{\boldsymbol{M}}_{\mathrm{Fe}}\right) \tag{2}$$

Where $\gamma$ is the gyromagnetic factor, $\alpha$ is the damping factor and $\boldsymbol{M}_{\mathrm{Tb}}$ and $\boldsymbol{M}_{\mathrm{Fe}}$ are the magnetization of Tb and FeCo sublattices. In the micromagnetic model, the effective field $\boldsymbol{H}_{\mathrm{eff}}$ is the sum of the external field $\boldsymbol{H}_{\mathrm{ext}}$, the demagnetization field $\boldsymbol{H}_{\mathrm{demag}}$, the anisotropy field $\boldsymbol{H}_{\mathrm{ani}}$, and the exchange field $\boldsymbol{H}_{\mathrm{exch}}$. In this model, each component has its respective effective field.

$$\boldsymbol{H}_{\mathrm{eff}_{\mathrm{Tb}}} = \boldsymbol{H}_{\mathrm{ext}_{\mathrm{Tb}}} + \boldsymbol{H}_{\mathrm{demag}_{\mathrm{Tb}}} + \boldsymbol{H}_{\mathrm{ani}_{\mathrm{Tb}}} + \boldsymbol{H}_{\mathrm{exch}_{\mathrm{Tb}}} \tag{3}$$

$$\boldsymbol{H}_{\mathrm{eff}_{\mathrm{Fe}}} = \boldsymbol{H}_{\mathrm{ext}_{\mathrm{Fe}}} + \boldsymbol{H}_{\mathrm{demag}_{\mathrm{Fe}}} + \boldsymbol{H}_{\mathrm{ani}_{\mathrm{Fe}}} + \boldsymbol{H}_{\mathrm{exch}_{\mathrm{Fe}}} \tag{4}$$

Here, we assume the effective external field and demagnetization field are equal for both sublattices in the same cell. However, the effective anisotropy field of the two sublattices are different due to their different anisotropy constants $K_{\mathrm{u}_{\mathrm{Tb}}}$ and $K_{\mathrm{u}_{\mathrm{Fe}}}$. Finally, since each sublattice interacts with itself and the other, the effective exchange field must contains contributions from interactions within the same sublattice, $\boldsymbol{H}_{\mathrm{exch}_{\mathrm{Tb-Tb}}}$ and $\boldsymbol{H}_{\mathrm{exch}_{\mathrm{Fe-Fe}}}$, and interactions between the sublattices $\boldsymbol{H}_{\mathrm{exch}_{\mathrm{Tb-Fe}}}$and $\boldsymbol{H}_{\mathrm{exch}_{\mathrm{Fe-Tb}}}$. The total effective exchange fields in the two-sublattice model can be expressed as the following equations.

$$\boldsymbol{H}_{\mathrm{exch}_{\mathrm{Tb}}} = \frac{2A_{\mathrm{Tb-Tb}}}{\mu_0 M_{\mathrm{Tb}}}\nabla^2\boldsymbol{m}_{\mathrm{Tb}} + \frac{2A_{\mathrm{Tb-Fe}}}{\mu_0 M_{\mathrm{Tb}}}\nabla^2\boldsymbol{m}_{\mathrm{Fe}} + \frac{B_{\mathrm{Tb-Fe}}}{\mu_0 M_{\mathrm{Tb}}}\boldsymbol{m}_{\mathrm{Fe}} \tag{5}$$

$$\boldsymbol{H}_{\text{exch}_{\text{Fe}}} = \frac{2A_{\text{Fe-Fe}}}{\mu_0 M_{\text{Fe}}} \nabla^2 \boldsymbol{m}_{\text{Fe}} + \frac{2A_{\text{Fe-Tb}}}{\mu_0 M_{\text{Fe}}} \nabla^2 \boldsymbol{m}_{\text{Tb}} + \frac{B_{\text{Fe-Tb}}}{\mu_0 M_{\text{Fe}}} \boldsymbol{m}_{\text{Tb}} \tag{6}$$

Where $A_{\text{Tb-Tb}}$, $A_{\text{Fe-Tb}}$, $A_{\text{Tb-Fe}}$ and $A_{\text{Fe-Fe}}$ are exchange stiffness constants, and $B_{\text{Tb-Fe}}$ and $B_{\text{Fe-Tb}}$ are exchange interaction constants between sublattices. The detailed derivations are included in Appendix A. With these effective field definitions we solve the LLG equations (1) and (2) by employing the finite distance method based on the mircomagnetic package OOMMF [21].

## 3. Result and Discussion

### *3.1 Experimental Result*

#### *3.1.1 Above the compensation temperature*

EB effect has been found to exist in TbFeCo films that contain two nanoscale amorphous phases [9]. As revealed by atom probe tomography, scanning transmission electron microscopy, and energy dispersive spectroscopy mapping, one phase (Phase II) corresponds to regions of FeCo enrichment and Tb depletion, while the other (Phase I) corresponds to regions of Tb enrichment and FeCo depletion. The length scales of these two phases are 2-5 nm. The exchange interaction between the two phases is believed to lead to the observed EB in these TbFeCo films. More specifically, the Fe-enriched Phase II is dominated by FeCo moments at room temperature, and behaves in a FM manner. On the other hand, the Tb-enriched Phase I is a near-compensated FiM with a large coercivity. When the field is not large enough to switch Phase I, Phase I provides unidirectional exchange anisotropy and affects the reversal field of Phase II. It should be noted that the EB effect in this system is a minor loop effect, arising from the fact that Phase I could only be switched in a sufficiently large field.

**Fig.** 1 shows the EB minor loops of TbFeCo above $T_{\text{comp}}$. Both positive and negative EB are observed at 300 K. Negative EB is observed in sample initialized at 355 K and 3 T, then, cooled down to 300 K at zero field. Hysteresis loop is measured at 300 K from 3 T through -3 T to 3 T. At 355 K and 3 T, the FeCo moments of both Phase I and Phase II are aligned in the positive direction, parallel to the applied field. Cooling down to 300 K at zero field maintains the spin orientation for both Phase I and Phase II. At 300 K, the near-compensated phase, Phase I, has larger coercivity then 3 T. Thus, within 3 T external field, Phase I maintains its spin orientation. Since the FeCo moments in both Phase I and Phase II are orientated parallel to each other at 3 T, additional external field is required to reverse the moments of Phase II when going from 3 T to -3 T, resulting in negative EB. The observed shift in overall magnetization originates from Phase I. Since the moments of Phase I maintain their orientations, they contribute a fixed amount to the overall magnetization of the sample. After initializing at 355K and 3T, Phase I has net positive magnetization at 300 K, resulting in a positive shift in the magnetization. Positive EB is observed in sample initialized at 175 K and 3 T, then, warmed up to 300 K at zero field. At 175 K and 3 T, since it is below $T_{\text{comp}}$ of Phase I,

the FeCo moments of Phase I are aligned in the negative direction, opposite to the applied field. This initializing procedure results in the FeCo moments of Phase I (negative) and Phase II (positive) aligned opposite to each other at 300 K and 3 T. Therefore, when the external field is applied from 3 T to -3 T, Phase I provides an additional energy to flip the moments of Phase II, resulting in a positive EB. The observed shift in overall magnetization is in the negative direction, because the net magnetization of Phase I is negative at 300 K after initializing at 175 K and 3 T.

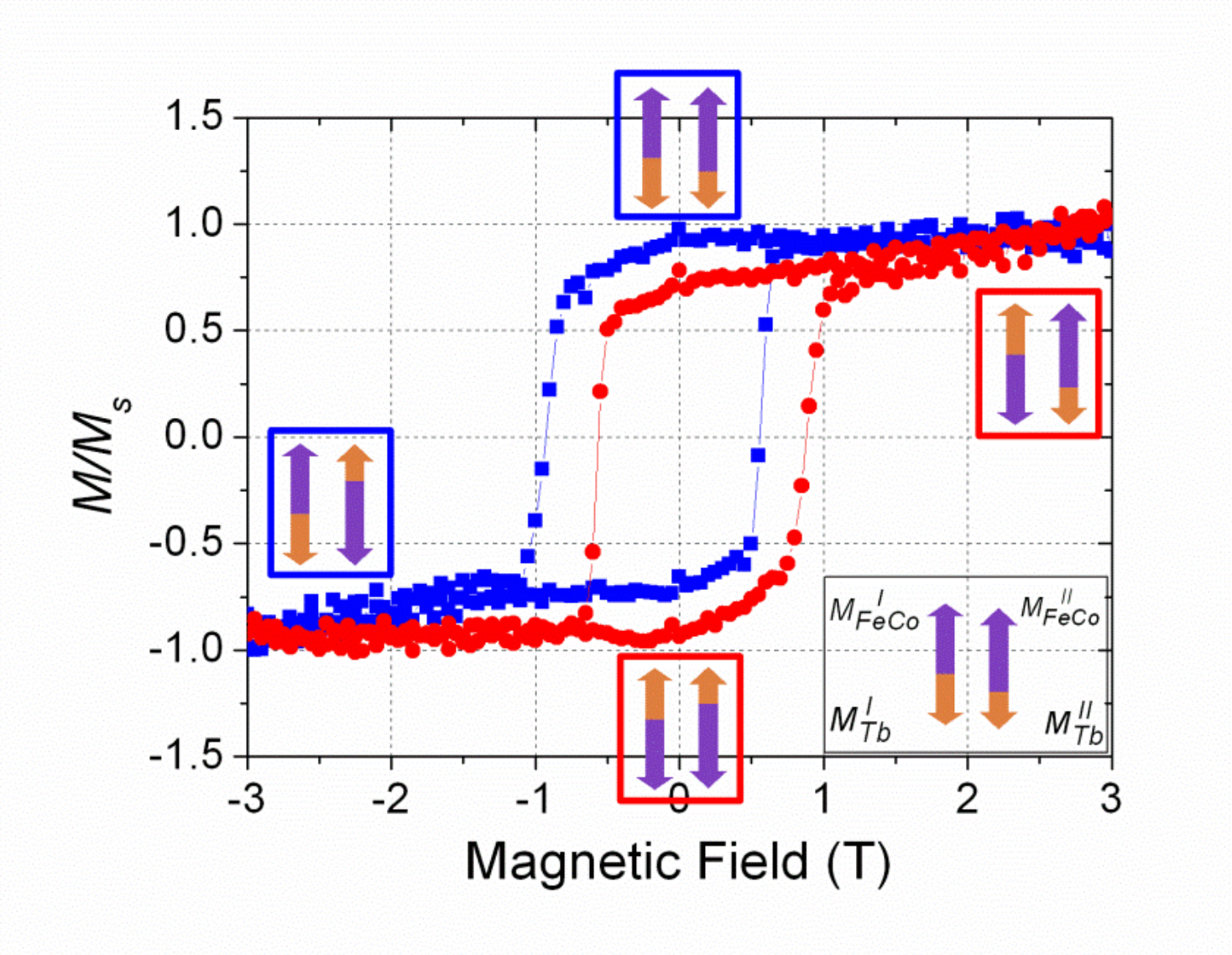

**Figure 1.** Experimental exchange bias minor loops of TbFeCo above $T_{comp}$. Sample initialized under 355 K and 3 T (blue square), and 175 K and 3 T (red circle). The insert shows an example of magnetic configuration. The left pair corresponds to the near-compensated Phase I, and the right for the uncompensated Phase II. Purple arrow represents the moments of FeCo, and orange arrow represents the moments of Tb. The blue box indicates the magnetic configuration of the sample initialized under 355K and 3T (blue square), and the red box indicates the magnetic configuration of the sample initialized under 175K and 3T (red circle).

### *3.1.2 Below the compensation temperature*

EB effect is also observed below $T_{comp}$. **Fig.** 2 shows hysteresis loops of TbSmFeCo below $T_{comp}$ of Phase I. At 300 K, applying external field from 3 T through -1 T to 3 T results in positive EB. Below $T_{comp}$ of Phase I, RE moments of Phase I dominate. At 300 K and 3 T, the FeCo moments of Phase I align in the negative direction, opposite to the applied field and the FeCo moments of Phase II. At -1 T, since coercivity of Phase I is larger than 1 T, the moments of Phase I maintain their orientations. On the other hand, the moments of Phase II are reversed, and align in the same direction as those of Phase I. Since it is favorable for moments of Phase I and Phase II to align in the same direction, a smaller external field is required to reverse the moments of Phase II when going from 3 T to -1 T, resulting in positive EB. Applying external field from -3 T through 1 T to -3 T results in negative EB. At -3 T, the FeCo moments of Phase I align in the positive direction, while those of Phase II align in the negative direction. At 1 T, only moments of Phase II are reversed, and align in the same direction as those of Phase I. Therefore, when the external field is applied from -3 T to 1 T, Phase I provides additional energy to flip the moments of Phase II, resulting negative EB.

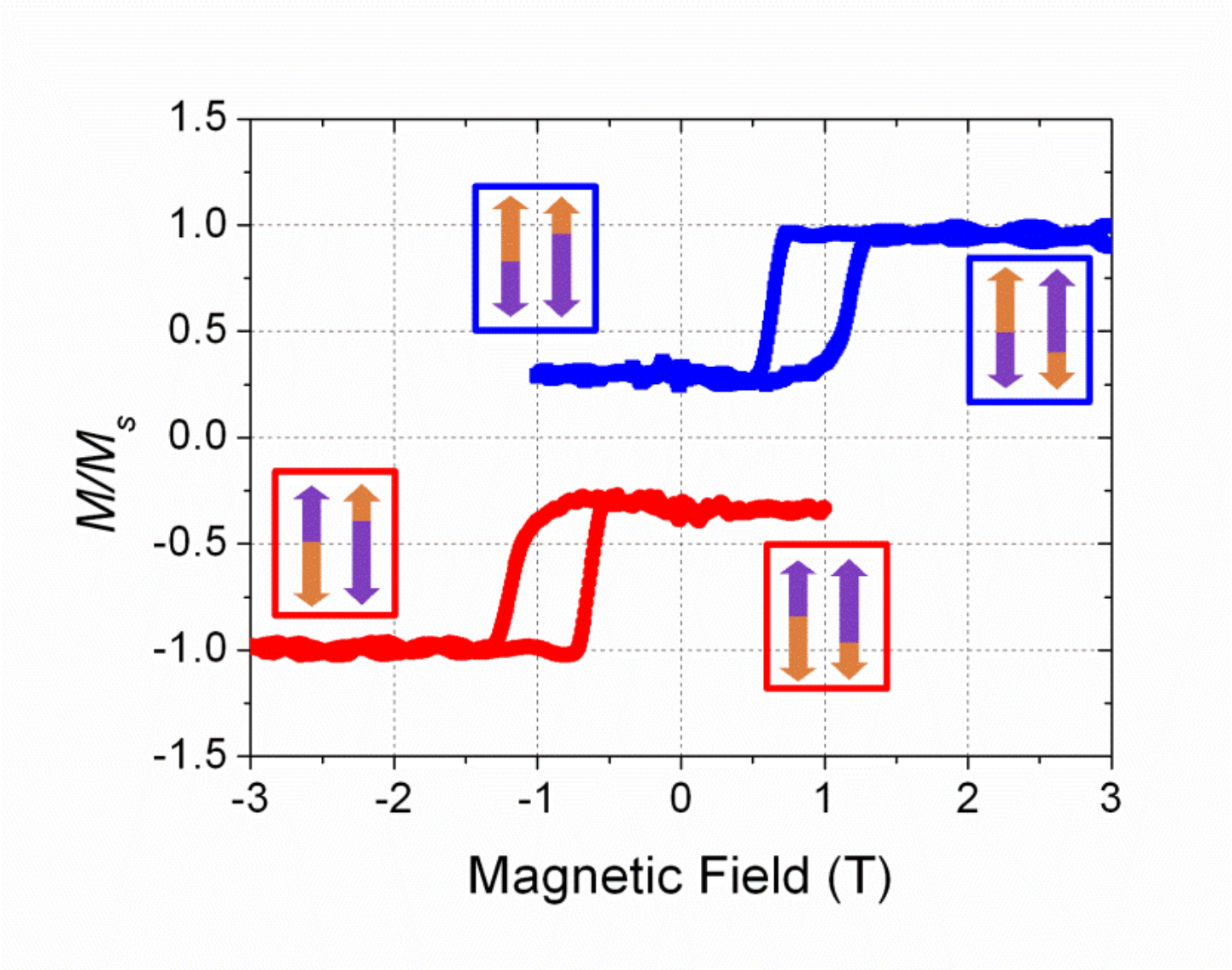

**Figure 2**. Experimental exchange bias minor loops of TbSmFeCo below $T_{\text{comp}}$ . External field scans from 3 T through -1 T to 3 T (blue solid), and from -3 T through 1 T to -3 T (red solid). The arrows are defined similarly as **Figure 1.**

### *3.2 Two-phase model*

In order to quantitatively validate the origin of the EB effect observed in the phase-separated amorphous TbFeCo films, a two-phase model is developed. Two kinds of cells, Tb-enriched and Fe-enriched, are used to represent the two nanoscale phases. Periodic boundary condition is employed. Various cell sizes, ranging from 0.5 nm to 1.0 nm, have been employed, and similar results are obtained. Only results of 0.5 nm × 0.5 nm × 0.5 nm cell are shown herein. Local Tb-enriched (or Fe-enriched) nanophase is modeled by a cubic block containing 216 Tb-enriched (or Fe-enriched) cells. Each block is 3-nm wide, comparable to the ~2-5-nm nanophases observed in experiment. There are 13 Phase I and 14 Phase II blocks to maintain right average composition. To capture the amorphous nature of the TbFeCo films, these blocks are distributed randomly in the cubic modeling space. It should be noted that in this simulation, there are two distinct compositions for each phase. In reality, there are variations in compositions for each phase, and the boundaries between the two phases are more gradual than the shape transitions employed in this simulation. The magnetic parameters of each type of cells are shown in **Table** 1. These parameters are derived from equations in Appendix A using exchange constants reported by Hansen et al. [8]. The anisotropy axis of each cell is distributed within a 45-degree cone, which is consistent with the amorphous nature of TbFeCo. Since we are only interested in static behavior, we set the effective damping constant $\alpha_{\text{eff}} = 1$. An external magnetic field $\boldsymbol{H}_{\text{ext}}$ is applied along the axis of the anisotropy cone to study the hysteresis loops of this two-phase system. **Fig.** 3 shows the temperature dependence of saturated magnetization ($M_s$) of simulated TbFeCo using the two-phase model, verifying $T_{\mathrm{comp}}$of the whole system is near 250 K, comparable to the experimental result. In the simulations, EB effects are observed both above and below $T_{\mathrm{comp}}$, and discussed in the following sections.

**Table 1.** Magnetic anisotropy constant and exchange constants of each type block used in the simulation.

| | Type 1 | Type 2 |
|---|---|---|
| $K_{\mathrm{Tb}}(\mathrm{J/m^3})$ | 3.4×10$^{5}$ | 1.9×10$^{5}$ |
| $A_{\mathrm{Tb-Tb}}$ (J/m) | 1.90×10$^{-12}$ | 1.21×10$^{-12}$ |
| $A_{\mathrm{Tb-Fe}}$ (J/m) | -2.43×10$^{-12}$ | -1.87×10$^{-12}$ |
| $A_{\mathrm{Fe-Fe}}$ (J/m) | 1.40×10$^{-11}$ | 1.68×10$^{-11}$ |
| $B_{\mathrm{Tb-Fe}}(\mathrm{J/m^3})$ | -1.43×10$^{7}$ | -1.09×10$^{7}$ |

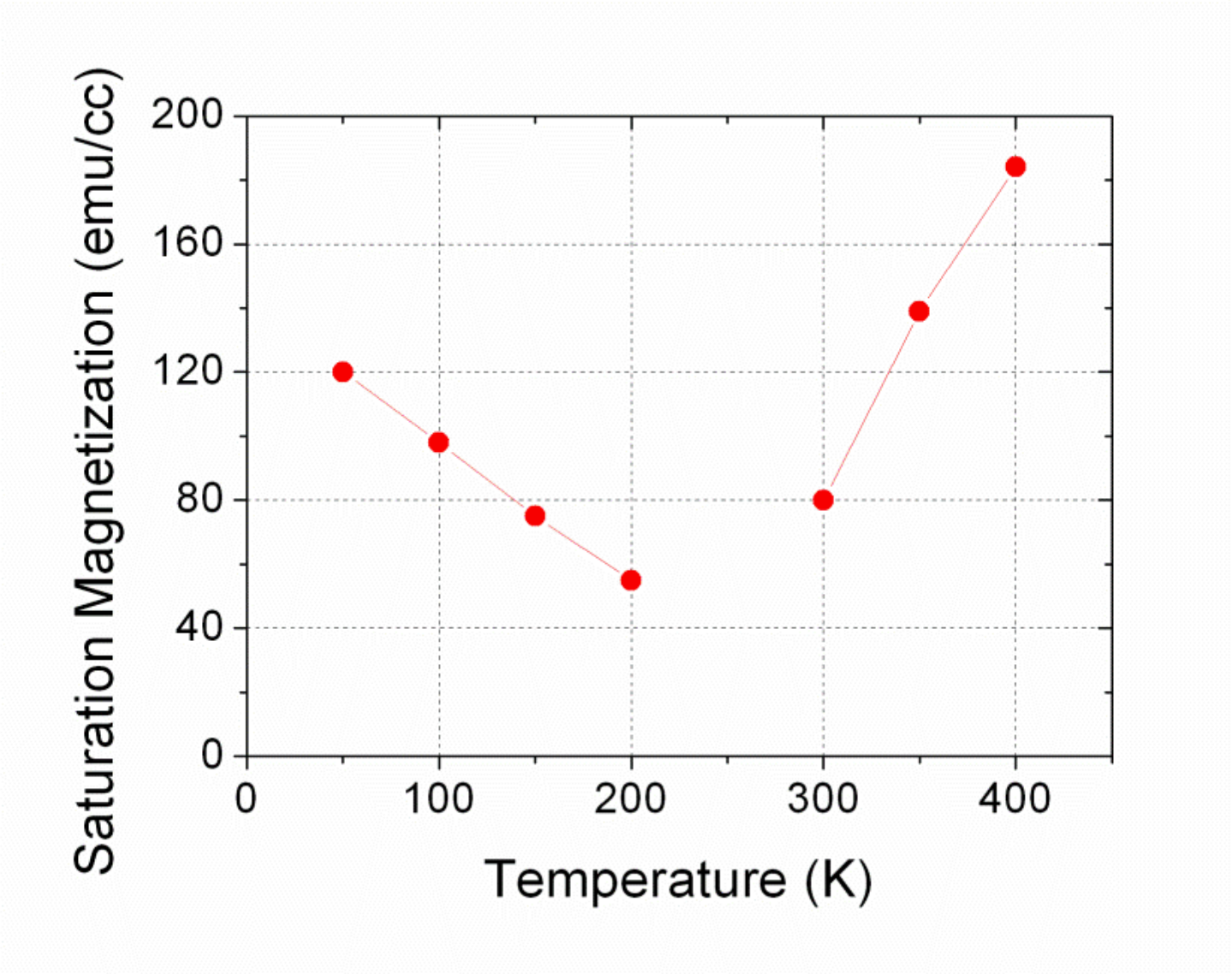

**Figure 3.** Temperature dependence of saturation magnetization simulated TbFeCo using the two-phase model.

### *3.2.1 Above the compensation temperature*

First, EB effect is observed above $T_{\mathrm{comp}}$. **Fig.** 4 shows the computed hysteresis loops at 300 K. In **Fig.** 4(a), with sufficient field, moments of Phase I and Phase II are able to reverse and result in a symmetric major loop. **Fig.** 4(b) and (c) show the contribution to the major loop from each phase. Clearly, Phase I has larger coercivity than Phase II. **Fig.** 4(d) shows EB minor loops above $T_{\mathrm{comp}}$. Applying external field from 5T through -1.1T to 5T results in negative EB minor loop. This is analogous to initialize the sample at 350 K and 3 T, then cool down to 300 K, and measure hysteresis loop in experiment. More specifically, at 5 T external field, the FeCo moments of both Phase I and Phase II are aligned in positive direction, in parallel to the external field, same as the spin configuration at 350 K and 3 T in experiment. At -1.1T external field, since Phase I has coercivity larger than 1.1 T, only the moments of Phase II are reversed. Similarly, in experiment, at 300 K, coercivity of Phase I is larger than 3 T, so only the moments of Phase II are reversed. Therefore, applying external field from 5 T through -1.1 T to 5 T results in negative EB minor loops, in agreement with experiment. Positive EB minor loop is observed by applying external field from -5T through 1.1T to -5T. This is analogous to initialize sample at 175 K and 3 T, then warm up to 300 K to measure hysteresis loop, resulting in positive EB. The shift in the hysteresis loops along the field axis ($|H_{\mathrm{E}}|$) is ~0.4 T. From **Fig.** 1, $|H_{\mathrm{E}}|$ is ~0.3 T in experiment. They are in excellent agreement. Using the same initial spin configurations as experiment, this two-phase model obtains both positive and negative EB minor loops, and $|H_{\mathrm{E}}|$ in agreement with experiment. Therefore, this two-phase model confirms that the exchange coupling between the two phases observed in TbFeCo is the origin of the EB effect in this system. In addition to this 3-nm two-phase model, smaller sizes of nanoscale phase separations have been used to investigate the limit of this EB effects. EB effects are observed in phase separation down to 1.5nm. However, due to the limit of micromagnetic model, where continuum approximation becomes questionable. Further numerical calculations using atomistic model are needed to determine the phase separation size of which EB effect vanishes.

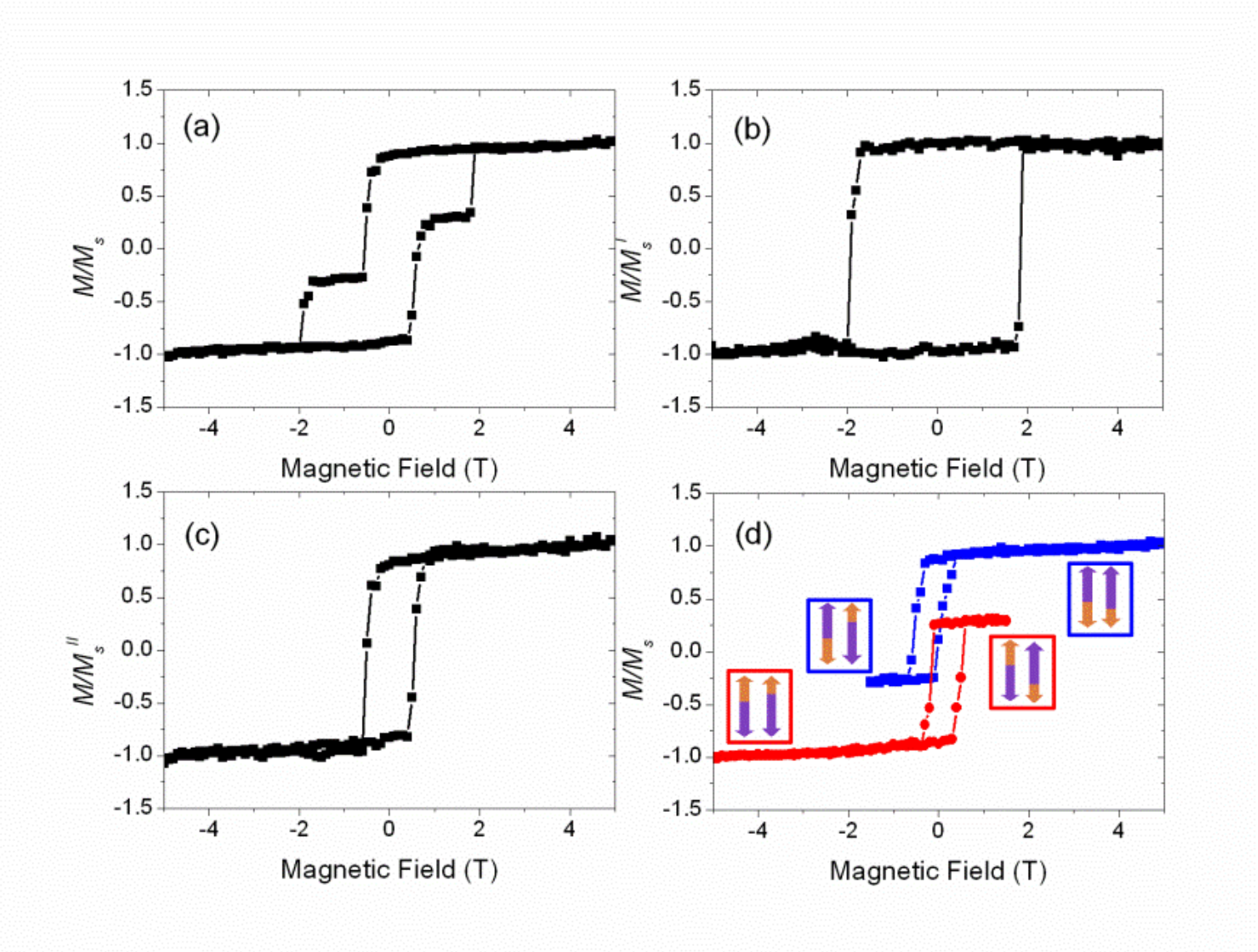

**Figure 4.** Simulated hysteresis loops of two-phase model. (a) Major loop of TbFeCo above $T_{comp}$, external field scans from 5 T to -5 T to 5 T. (b-c) Contribution to the major loop from Phase I (b) and Phase II (c) above $T_{\mathrm{comp}}$. (d) Exchange bias minor loops of TbFeCo above $T_{\mathrm{comp}}$. External field scans from 5 T to -1.1 T to 5 T (blue square), and from -5 T to 1.1 T to 5 T (red circle). The arrows are defined similarly as **Figure 1.**

### *3.2.2 Below the compensation temperature*

EB effect is also observed below $T_{\mathrm{comp}}$. **Fig.** 5 shows the EB minor loops of TbFeCo at 200 K, below $T_{\mathrm{comp}}$of Phase I. Both positive and negative EB is observed. Positive EB is obtained when external field is applied from 5T through -3T to 5T. On the other hand, negative EB is obtained when external field is applied from -5T through 3T to -5T. Compare to EB effect above $T_{\mathrm{comp}}$, the signs of EB correspond to opposite initial spin configurations. This is due to the fact that Phase I is below $T_{\mathrm{comp}}$, but Phase II is above $T_{\mathrm{comp}}$. Since Phase I is below $T_{\mathrm{comp}}$, the Tb moments dominate. In sufficiently high field, the FeCo moments of Phase I align antiparallel to the external field while the FeCo moments of Phase II align parallel to the external field. As a result, Phase I provides additional exchange anisotropic energy to favor the magnetic reversal of Phase II going from 5 T to -3 T, but introduces additional barrier going from -3 T to 5 T, resulting in positive EB effect. Negative EB effect can be understood similarly. $|H_E|$ is ~1.4 T, compared to the experimental value ~0.9T as shown in **Fig.** 2. The difference in $|H_E|$ is due to the fact that for simplicity, Sm has not been taken into account in the numerical calculation. Since Sm has low Neel temperature, it is approximately grouped into the same sublattice as Tb in the two-phase model. Comparing to TbSmFeCo, TbFeCo has larger perpendicular magnetic anisotropy, result in a larger coercivity seen in simulation. Therefore, with this micromagnetic model, EB effect is obtained for both above and below $T_{\mathrm{comp}}$.

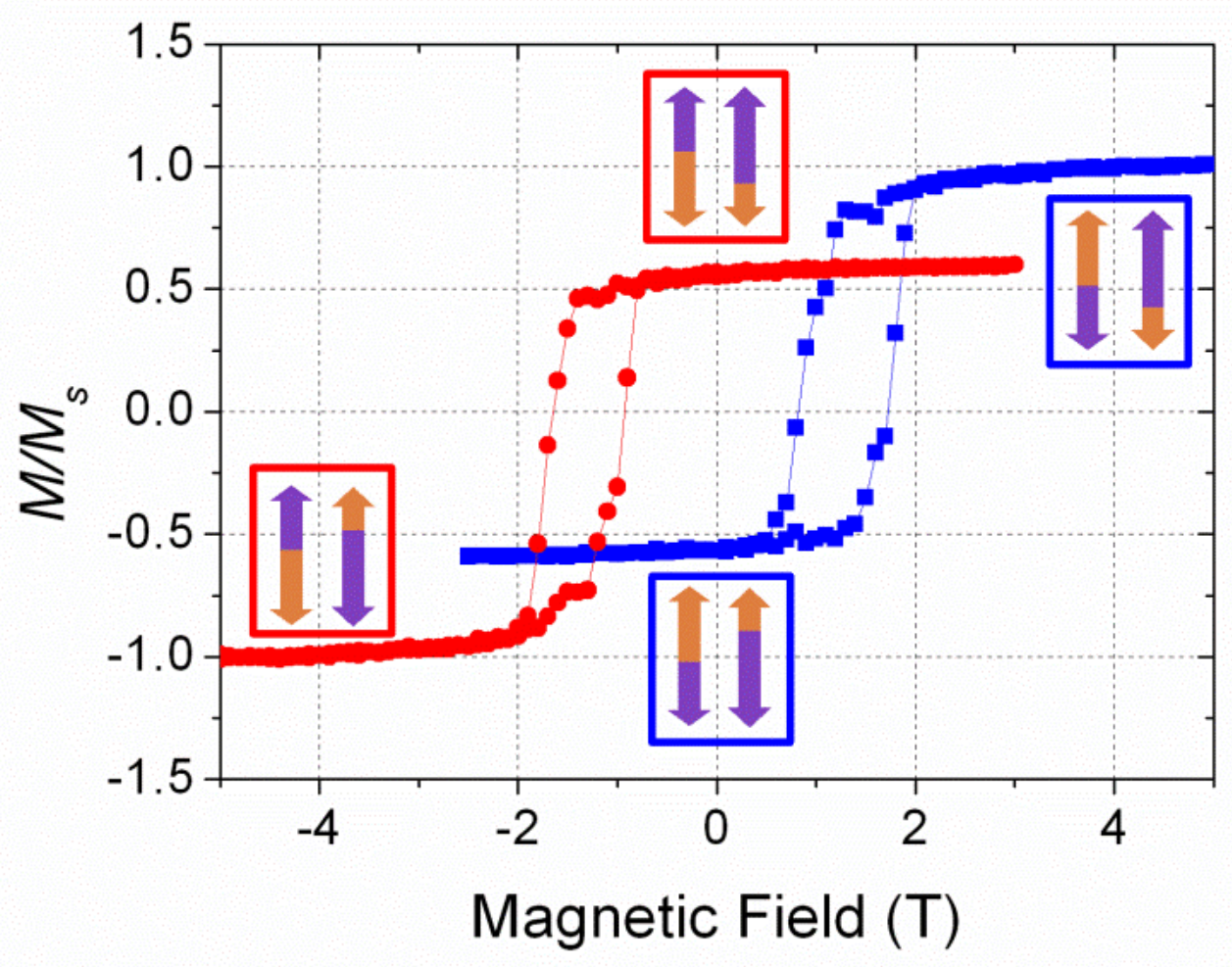

**Figure 5**. Simulated exchange bias minor loops of TbFeCo below $T_{\mathrm{comp}}$. External field scans from 5 T through -3 T to 5 T (blue square), and from -5 T through 3T to -5 T (red circle). The arrows are defined similarly as **Figure 1.**

### *3.2.3 Applications of the two-phase model in other systems*

In addition to understand the EB effect in a two-phase RE-TM system, this mircomagnetic model can also be employed to study EB effects in other two-phase materials with FiM phase. For example, intrinsic EB effects have been reported in polycrystalline Heusler alloys at low temperature [22-25]. An exchange field of more than 3T is uncovered in Mn-Pt-Ga with coexistence of FM and FiM regions, and shows strong dependence on compositions and field-cooled procedures [25]. The two-phase model can be employed to study this tunable EB effect in Mn-Pt-Ga. With different compositions of the FiM phase and initialization conditions, one can understand how they contribute to the tunable EB effect and lead to the development of new two-phase EB materials using FiM to achieve desirable properties for applications.

## 4. Summary

Micromagnetic model is employed to simulate the EB effect in FiM TbFeCo films containing two nanoscale phases. The original model is extended to allow the magnetic moments of each sublattice to evolve individually. Two types of cells and blocks with distinct Tb concentrations are developed in order to incorporate the two nanoscale phases. 8 Phase I blocks and 19 Phase II blocks are randomly distributed in a $3 \times 3 \times 3$ cube to model a structure with the two nanoscale phases. This model verifies that the observed EB effect in this FiM TbFeCo films originates from the exchange interaction between the two nanoscale phases. Moveover, both positive and negative EB loops have been observed above and below $T_{\mathrm{comp}}$, and the signs of EB effect are in agreement with the experimental results. Using this micromagnetic model, one can explore FiM/FM and FiM/FiM systems by tuning the composition of the nearly compensated FiM phase, and develop desirable EB properties for applications at room temperature.

### Acknowledgements

The work was partially supported by the Defense Threat Reduction Agency grant (Award No. HDTRA 1-11-1-0024).

## Appendix A

*Derivation of effective field due to exchange interaction in the two-sublattice model*

The Hamiltonian of nearest neighbor exchange interaction between site i and site j is

$$\mathcal{H}_A = -\frac{1}{2}\sum_{<i,j>} J_{ij}\,\boldsymbol{S}_i \cdot \boldsymbol{S}_j$$

$$= -\frac{1}{2}\sum_{<Tb_i,Tb_j>} J_{Tb-Tb}\boldsymbol{S}_{Tb_i}\cdot\boldsymbol{S}_{Tb_j} - \frac{1}{2}\sum_{<Fe_i,Fe_j>} J_{Fe-Fe}\boldsymbol{S}_{Fe_i}\cdot\boldsymbol{S}_{Fe_i} - \sum_{<Tb_i,Fe_j>} J_{Tb-Fe}\boldsymbol{S}_{Tb_i}\cdot\boldsymbol{S}_{Fe_j}$$

Where $\boldsymbol{S_A}$ is the moment of element A.

We can rewrite Tb-Tb and Fe-Fe terms as follow

$$\mathcal{H}_{Tb-Tb} = -\frac{1}{2}J_{Tb-Tb}S_{Tb}^2\sum_{<Tb_i,Tb_j>}\boldsymbol{m}_{Tb_i}\cdot\boldsymbol{m}_{Tb_j}$$

$$= const. + \frac{1}{4}J_{Tb-Tb}S_{Tb}^2\sum_{<Tb_i,Tb_j>}\left(\boldsymbol{m}_{Tb_i}-\boldsymbol{m}_{Tb_j}\right)^2$$

Using the continuous assumption

$$\boldsymbol{m}_{Tb_j} \approx \boldsymbol{m}_{Tb_i} + \boldsymbol{r}_{ij}\cdot\nabla\boldsymbol{m}_{Tb_i}$$

$$\mathcal{H}_{Tb-Tb} \approx \frac{1}{4}J_{Tb-Tb}S_{Tb}^2 z_{Tb-Tb} r_{nn}^2\sum_{Tb_i}\left(\nabla\boldsymbol{m}_{Tb_i}\right)^2 = A_{Tb-Tb}\int(\nabla\boldsymbol{m}_{Tb})^2 d^3x$$

Similarly,

$$\mathcal{H}_{Fe-Fe} \approx \frac{1}{4}J_{Fe-Fe}S_{Fe}^2 z_{Fe-Fe} r_{nn}^2\sum_{Fe_i}\left(\nabla\boldsymbol{m}_{Fe_i}\right)^2 = A_{Fe-Fe}\int(\nabla\boldsymbol{m}_{Fe})^2 d^3x$$

$A_{Tb-Tb} = \frac{1}{4}J_{Tb-Tb}S_{Tb}^2 z_{Tb-Tb}c_{Tb}/r_{nn}$ and $A_{Fe-Fe} = \frac{1}{4}J_{Fe-Fe}S_{Fe}^2 z_{Fe-Fe}c_{Fe}/r_{nn}$

Where $c_A$is the element A concentration, $z_{A-B}$is the number of element B atoms around element A, and $r_{nn}$is the distance to the nearest neighbor.

The ferrimagnetic (Tb-Fe) term

$$\mathcal{H}_{Tb-Fe} = -\sum_{<Tb_i,Fe_j>} J_{Tb-Fe}\boldsymbol{S}_{Tb_i}\cdot\boldsymbol{S}_{Fe_j} = \frac{1}{2}J_{Tb-Fe}S_{Tb}S_{Fe}\sum_{<Tb_i,Fe_j>}\left(\boldsymbol{m}_{Tb_i}-\boldsymbol{m}_{Fe_j}\right)^2$$

Using the continuous assumption to expand $\boldsymbol{m}_{Fe_j}$

$$\mathcal{H}_{Tb-Fe} \approx \frac{1}{2} J_{Tb-Fe} S_{Tb} S_{Fe} \sum_{<Tb_i, Fe_j>} \left( \boldsymbol{m}_{Tb_i} - \boldsymbol{m}_{Fe_i} - \boldsymbol{r}_{ij} \cdot \nabla \boldsymbol{m}_{Fe_i} - \frac{1}{2} \boldsymbol{r}_{ij}^2 \nabla^2 \boldsymbol{m}_{Fe_i} \right)^2$$

$$\approx \frac{1}{2} J_{Tb-Fe} S_{Tb} S_{Fe} \sum_{<Tb_i, Fe_j>} \Big( (\boldsymbol{m}_{Tb_i} - \boldsymbol{m}_{Fe_i})^2 - 2(\boldsymbol{m}_{Tb_i} - \boldsymbol{m}_{Fe_i}) \cdot (\boldsymbol{r}_{ij} \cdot \nabla \boldsymbol{m}_{Fe_i}) - (\boldsymbol{m}_{Tb_i} - \boldsymbol{m}_{Fe_i}) \boldsymbol{r}_{ij}^2 \cdot \nabla^2 \boldsymbol{m}_{Fe_i} + (\boldsymbol{r}_{ij} \cdot \nabla \boldsymbol{m}_{Fe_i})^2 \Big)$$

The second term $\sum_{<Tb_i, Fe_j>} \left( -2(\boldsymbol{m}_{Tb_i} - \boldsymbol{m}_{Fe_i}) \cdot (\boldsymbol{r}_{ij} \cdot \nabla \boldsymbol{m}_{Fe_i}) \right)$ vanishes with the assumption of center symmetry

Combine the last two terms,

$$\mathcal{H}_{Tb-Fe} \approx \frac{1}{2} J_{Tb-Fe} S_{Tb} S_{Fe} z_{Tb-Fe} \sum_{Tb_i} \left( (\boldsymbol{m}_{Tb_i} - \boldsymbol{m}_{Fe_i})^2 - r_{nn}^2 \boldsymbol{m}_{Tb_i} \cdot \nabla^2 \boldsymbol{m}_{Fe_i} + r_{nn}^2 \nabla \cdot (\boldsymbol{m}_{Fe_i} \cdot \nabla \boldsymbol{m}_{Fe_i}) \right)$$

$$= -B_{Tb-Fe} \int \boldsymbol{m}_{Tb} \cdot \boldsymbol{m}_{Fe} d^3x - 2A_{Tb-Fe} \int \boldsymbol{m}_{Tb} \cdot \nabla^2 \boldsymbol{m}_{Fe} d^3x + 2A_{Tb-Fe} \oint \boldsymbol{m}_{Fe} \cdot \nabla \boldsymbol{m}_{Fe} \cdot \boldsymbol{n} dS$$

$$A_{Tb-Fe} = \frac{1}{4} J_{Tb-Fe} S_{Tb} S_{Fe} z_{Tb-Fe} c_{Tb} / r_{nn}$$

$$A_{Fe-Tb} = \frac{1}{4} J_{Tb-Fe} S_{Tb} S_{Fe} z_{Fe-Tb} c_{Fe} / r_{nn}$$

and $B_{Tb-Fe} = J_{Tb-Fe} S_{Tb} S_{Fe} c_{Tb} z_{Tb-Fe} / a^3 = B_{Fe-Tb}$

The total energy

$$\mathcal{H}_A = \int \left( A_{Fe-Fe} (\nabla \boldsymbol{m}_{Fe})^2 + A_{Tb-Tb} (\nabla \boldsymbol{m}_{Tb})^2 - 2A_{Tb-Fe} \boldsymbol{m}_{Tb} \cdot \nabla^2 \boldsymbol{m}_{Fe} - B_{Tb-Fe} (\boldsymbol{m}_{Tb} \cdot \boldsymbol{m}_{Fe}) \right) d^3x + 2A_{Tb-Fe} \oint \boldsymbol{m}_{Fe} \nabla \boldsymbol{m}_{Fe} \cdot \boldsymbol{n} dS$$

The last term is integrated on the boundary, so the energy density is

$$\mathcal{E}_A = A_{Fe-Fe} (\nabla \boldsymbol{m}_{Fe})^2 + A_{Tb-Tb} (\nabla \boldsymbol{m}_{Tb})^2 - 2A_{Tb-Fe} \boldsymbol{m}_{Tb} \nabla^2 \boldsymbol{m}_{Fe} - B_{Tb-Fe} (\boldsymbol{m}_{Tb} \cdot \boldsymbol{m}_{Fe})$$

The effective field due to exchange interaction

$$\boldsymbol{H}_{eff,Tb} = -\frac{\delta \mathcal{E}_A}{\mu_0 M_{s,Tb} \delta \boldsymbol{m}_{Tb}}$$

$$= \frac{2}{\mu_0 M_{s,Tb}} A_{Tb-Tb} \nabla^2 \boldsymbol{m}_{Tb} + \frac{2}{\mu_0 M_{s,Tb}} A_{Tb-Fe} \nabla^2 \boldsymbol{m}_{Fe} + \frac{1}{\mu_0 M_{s,Tb}} B_{Tb-Fe} \boldsymbol{m}_{Fe}$$

Similarly,

$$\boldsymbol{H}_{eff,Fe} = -\frac{\delta \mathcal{E}_A}{\mu_0 M_{s,Fe} \delta \boldsymbol{m}_{Fe}}$$

$$= \frac{2}{\mu_0 M_{s,Fe}} A_{Fe-Fe} \nabla^2 \boldsymbol{m}_{Fe} + \frac{2}{\mu_0 M_{s,Fe}} A_{Fe-Tb} \nabla^2 \boldsymbol{m}_{Tb} + \frac{1}{\mu_0 M_{s,Fe}} B_{Fe-Tb} \boldsymbol{m}_{Tb}$$